\documentstyle[12pt]{article}

\begin{document}

\begin{titlepage}

\hspace{9.5cm}{IFT-P.013/2001}

\vspace{.5cm}

\begin{center}

\LARGE

{\sc Review of Open Superstring Field Theory}

\vspace{.5cm}
\large

Nathan Berkovits\footnote{e-mail:  nberkovi@ift.unesp.br}

\vspace{.5cm}

{\em Instituto de F\'\i sica Te\'orica, Universidade Estadual Paulista} \\
{\em Rua Pamplona 145, 01405-900, S\~ao Paulo, SP, Brasil}

\vspace{.5cm}

May 2001

\end{center}

\vspace{1cm}

\begin{abstract}

I review the construction of an action for open superstring field
theory which does not suffer from the contact term problems of other
approaches. This action resembles a Wess-Zumino-Witten action
and can be constructed in a manifestly D=4 super-Poincar\'e
covariant manner. 
This review is based on lectures given at the ICTP Latin-American
String School in Mexico City and the Komaba 2000 Workshop in Tokyo.

\end{abstract}

\end{titlepage}

\newpage
\def\half{{1 \over 2}}
\def\dzm{{\partial_z}}
\def\dz{{\partial_z}}
\def\dzp{{\partial _{\bar z}}}
\def\dzbar{{\bar\partial _{\bar z}}}
\def\jb{{\bar j}}
\def\p{{\partial}}
\def\pj{{\partial_j}}
\def\pjb{{\bar\partial^j}}
\def\pk{{\partial_k}}
\def\pkb{{\bar\partial^\k}}
\def\N{{\nabla}}
\def\Nb{{\bar\nabla}}
\def\pb{{\bar p}}
\def\L{{\Lambda}}
\def\Gf{{{\cal{G}}_1^+}}
\def\Gs{{{\cal{G}}_0^+}}
\def\Gtf{{\tilde{\cal{G}}_{-2}^+}}
\def\Gts{{\tilde{\cal{G}}_{-1}^+}}
\def\P{{\Phi}}
\def\sm {{\psi^-}}
\def\sp {{\psi^+}}
\def\eps {{\epsilon}}
\def\ep {{\epsilon^{jk}}}
\def\epb {{\epsilon^{\bar j\bar k}}}
\def\xj {{x_j}}
\def\xk {{x_k}}
\def\xbj {{\bar x_{\bar j}}}
\def\xbk {{\bar x_{\bar k}}}
\def\pbk {{\psi^-_{\bar k}}}
\def\pk {{\psi^+_k}}
\def\a {{\alpha}}
\def\b {{\beta}}
\def\g {{\gamma}}
\def\d {{\delta}}
\def\e {{\epsilon}}
\def\E {{\epsilon}}
\def\eb {{\epsilon_{\bar\a\bar\b\bar\g\bar\d}}}
\def\eu {{\epsilon^{\a\b\g\d}}}
\def\ad {{\dot\alpha}}
\def\bd {{\dot\beta}}
\def\k {{\kappa}}
\def\kb {{\bar\kappa}}
\def\s {{\sigma}}
\def\t {{\theta}}
\def\ta {{\theta^\alpha}}
\def\pbad {{\bar p_{\ad}}}
\def\tba {{\bar\theta^\ad}}
\def\oj{{\omega^j}}
\def\obj{{\bar\omega_j}}
\def\ok{{\omega^k}}
\def\obk{{\bar\omega_k}}
\def\ol{{\omega^l}}
\def\obl{{\bar\omega_l}}
\def\O{{\Omega_{-1}}}
\def\Ob{{\Omega_0}}
\def\dj{{\partial^j}}
\def\djb{{\partial_{\bar j}}}
\def \ad {{\dot \a}}
\def \bd {{\dot \b}}
\def \t {{\theta}}
\def \tb {{\bar\theta}}
\def \Gp {{G^+}}
\def \Gtp {{\tilde G^+}}
\def \Gtm {{\tilde G^-}}
\def \Gpf {{G^+_1}}
\def \Gtpf {{\tilde G^+_{-2}}}
\def \Gps {{G^+_0}}
\def \Gtps {{\tilde G^+_{-1}}}

\section {Problems with Conventional Approach}

The construction of a field theory action for the superstring is
an important problem since it may lead to information about
non-perturbative superstring theory which is unobtainable from the
on-shell perturbative S-matrix.
This information might be useful for understanding the non-perturbative
dualities of the superstring. Although there was much activity ten years
ago concerning a field theory action for the bosonic string, there
was not much progress on constructing a field theory action for the
superstring. 

After discussing the problems with conventional approaches to
superstring field theory in section 1, a Wess-Zumino-Witten-like action
will be constructed for open Neveu-Schwarz string field theory in section 2.
In section 3, this action will be generalized to any open string with
critical N=2 superconformal invariance, and section 4 will review an
open superstring field theory action
with manifest four-dimensional
super-Poincar\'e covariance which
includes all sectors of the superstring.

The covariant string field theory action for the bosonic string is based on
a BRST operator $Q$ and a string field $V$ of $+1$ ghost-number. 
In Witten's approach to
open string field theory, the gauge-invariant action is
\cite{witbos}
\begin{equation}
S={1\over{\lambda^2}}
Tr \langle \half V Q V + {1\over 3} V^3 \rangle
\label{openbos}
\end{equation}
where string fields are glued together at their
midpoint. 

In generalizing this approach to superstring field theory, the
main difficulty comes from the requirement that the string field carries
a definite ``picture''. Recall that each physical state of the superstring
is represented by an infinite number of BRST-invariant vertex operators
in the covariant RNS formalism \cite{fms}. 
To remove this infinite degeneracy, one
needs to require that the vertex operator carries a definite picture,
identifying which modes of the $(\beta,\gamma)$ ghosts annihilate
the vertex operator.
For open superstring fields,
the most common choice is that all Neveu-Schwarz (NS) string fields carry
picture $-1$ and all Ramond (R) string fields carry picture $-\half$.

Since the total picture must equal $-2$ for open superstrings, the
obvious generalization of the action (\ref{openbos})
is
\cite{wsup}
\begin{equation}
S={1\over{\lambda^2}}
Tr \langle \half V_{NS}
Q V_{NS} +\half V_R Q Y V_R  + {1\over 3}
 Z ~V^3_{NS} +\half V_{NS} V_R V_R \rangle
\label{opensuper}
\end{equation}
where the $(\b,\g)$ ghosts are fermionized as
$\b=e^{-\phi}\p\xi$ and $\g= \eta e^\phi$,
$Z=\{Q,\xi\}$ is the picture-raising operator of picture $+1$,
$Y=c\p\xi e^{-2\phi}$ is the picture-lowering operator of picture $-1$,
and these picture-changing operators are inserted at the midpoint of
the interacting strings.
However, as shown by Wendt
\cite{conw},
the action of (\ref{opensuper}) is
not gauge-invariant because of the contact-term divergences occuring when two
$Z$'s collide. 
One way to make the action gauge-invariant
would be to introduce contact
terms to cancel the divergences coming from colliding $Z$'s.
However, the coefficients of these contact terms would have to
be infinite in the classical action since the divergences are present
already in tree-level amplitudes.
Note that infinite contact terms are also expected
in light-cone superstring field theory (either in the RNS or
Green-Schwarz formalisms) to cancel the divergences when
interaction points collide \cite{GK}.

Although one can choose other pictures for the string
field $V$ which change the relative factors of $Z$ and $Y$
\cite{pr}\cite{arone},
there is
no choice for which the action is cubic and gauge-invariant
\cite{bigp}. 
For example, choosing $V_{NS}$ in the zero
picture leads to the action
\cite{aref}
\footnote{This action was recently used to compute 
the tachyon potential in  NS string theory. 
However, besides the gauge invariance
problems mentioned here, there appears to be an error in their 
computation of the D-brane tension by a factor of $\sqrt{2}$.
When written
in terms of the closed string coupling constant, the tension
is background-dependent and picks up a factor of $\sqrt{2}$
if the D-brane is 
non-BPS. But when written in terms of the open string coupling
constant, the tension is background independent and does not
pick up a factor of $\sqrt{2}$
\cite{senuniv}\cite{zw}\cite{senpriv}\cite{ohm}.} 

\begin{equation}
S= {1\over{\lambda^2}}
Tr \langle\half V_{NS} Y^2 
Q V_{NS} + \half V_R Q Y V_R  + {1\over 3}
Y^2 ~V^3_{NS} + \half Y V_{NS} V_R V_R \rangle.
\label{arefaction}
\end{equation}
The kinetic term
of (\ref{arefaction}) 
implies that $Y^2 Q V_{NS} =0$
is the linearized equation of motion for the Neveu-Schwarz
field. But since $Y^2$ has a non-trivial kernel, this equation of motion has
additional solutions which are not in the cohomology of $Q$.
Although one could restrict
$V_{NS}$ to only include states not in
the kernel of $Y^2$, such a projection would break 
gauge-invariance
since Witten's midpoint gluing prescription does not preserve
this projection of the string field. Modifying the gluing prescription
to preserve the projection would ruin its associativity properties.
Note that a similar problem \cite{anom}
exists for the Ramond kinetic term in the
action of (\ref{opensuper}).

\section{ Open Neveu-Schwarz String Field Theory} 

In this section, it will be shown how to construct a ten-dimensional
Lorentz-covariant action for the Neveu-Schwarz sector of open
superstring field theory. It is not yet know how to extend this action
to the Ramond sector in a ten-dimensional
Lorentz-covariant manner. However, as will
be shown in section 4, it can be generalized to a four-dimensional
super-Poincar\'e
covariant action which includes both the NS and R sectors.

To construct a field theory action, one first needs to define a
NS string field. The first attempts to construct an action used the
fermionic string field
$$V=  c e^{-\phi} \psi^\mu A_\mu(x) + ...$$
where $V$ was constrained to carry $+1$ ghost number\footnote{
The ghost-number operator will be defined as $\int(cb+\eta\xi)$
so that $(\eta,\xi)$ carries ghost-number $(+1,-1)$ and $e^{n\phi}$
carries zero ghost number.}
and $-1$ picture,
and to satisfy $\eta_0 V=0$, i.e. to be independent
of the $\xi_0$ mode.
Although quadratic actions were succesfully constructed using $V$,
the cubic
interaction term had problems due to the necessity of introducing
the picture-raising operator $Z$.

The solution to this problem is to express $V$ and $ZV$ in terms
of a more fundamental NS string field $\Phi$ which is bosonic
and carries zero ghost-number and zero picture.
If one defines 
$\Phi= \xi_0 V$, i.e.
$$\Phi = \xi c e^{-\phi} \psi^\mu A_\mu(x) + ... , $$
then 
$V=\eta_0 \Phi$ and  
$ZV=Q\Phi$.
However, as will be shown below, $\Phi=\xi_0 V$ is a specific gauge
choice for $\Phi$, and one needs a more gauge-invariant description
for $\Phi$ in order to construct an NS string field theory action without
contact term problems.

The quadratic action for $\Phi$ will be defined as
$$S={1\over{2\lambda^2}}
Tr\langle \Phi ~Q \eta_0 ~\Phi\rangle.$$
Because of the $\xi$ zero mode, the non-vanishing inner product will
be defined as in the ``large'' Hilbert space of \cite{fms}, i.e. 
$$\langle \xi c \p c\p^2 c e^{-2\phi} \rangle =1.$$ 
Since
$Q^2 = \eta_0^2= \{Q,\eta_0\}=0$, $S$ is invariant under the linearized gauge
transformation
$$\d\Phi = \eta_0 \tilde\Lambda + Q \Lambda$$
where $\tilde\Lambda$ and $\Lambda$ are independent gauge parameters.
Note that $\Phi$ carries picture 0, $\eta_0$ carries picture $-1$ and
$Q$ carries picture 0, so $\tilde\Lambda$ and $\Lambda$ must
carry picture $+1$ and $0$ respectively.

Since $\{\eta_0,\xi_0\}=1$, the $\tilde\Lambda$ parameter can be used
to gauge $\Phi=\xi_0 V$ for some $V$ annihilated by $\eta_0$. 
The equation of motion
for $S$ is
$$Q\eta_0\Phi=0,$$
which in this gauge implies that
$$Q\eta_0 (\xi_0 V)= QV =0.$$
Furthermore, the remaining gauge parameter $\Lambda$ generates
the gauge transformation
$$\d V = \eta_0 (Q\Lambda) = Q\Omega$$
if one chooses $\Lambda=\xi_0\Omega.$ So one recovers the desired
linearized equations of motion and gauge invariances for $V$.

To include interactions, one needs to find an action which allows
a non-linear generalization of the gauge invariances
$\d\Phi = \eta_0 \tilde\Lambda + Q \Lambda$. This can be obtained
by drawing an analogy with the two-dimensional
Wess-Zumino-Witten (WZW) action \cite{nonab}
$$S_{WZW}= {1\over{2\lambda^2}}
Tr \int d^2 z ((g^{-1}\p g)(g^{-1}\bar\p g)
- \int_0^1 dt (\hat g^{-1}\p_t \hat g)
[ \hat g^{-1}\p \hat g ~,~\hat g^{-1}\bar\p \hat g]) $$
where $g(z,\bar z)$ is a group-valued two-dimensional field and
$\hat g(t,z,\bar z)$ is any continuous
group-valued three-dimensional
field defined on the three-volume with boundary at
$t=0$ and $t=1$ such that
$\hat g(1,z,\bar z)= g(z,\bar z)$
and $\hat g(0,z,\bar z)= 1$.
Recall that $S_{WZW}$ is invariant under the gauge transformation
$$g(z,\bar z) \to \bar\Omega(\bar z) g(z,\bar z) + g(z,\bar z) \Omega(z)$$
where $\p \bar\Omega(\bar z)=\bar\p \Omega(z)=0$.
If one writes $g= e^\Phi$ where $\Phi$ is Lie-algebra valued,
the gauge transformation on $\Phi$ is
$\d\Phi = \bar\Omega(\bar z) + \Omega (z) + ... $
where $...$ depends on $\Phi$.
Furthermore, the WZW equation of motion $\bar\p (g^{-1} \p g)=0$
implies that
$\p\bar\p\Phi= ...$
where $...$ is non-linear in $\Phi$.

This suggests writing the NS string field theory action as
\begin{equation}
S= {1\over{2\lambda^2}}
 Tr \langle (e^{-\Phi}Q e^\Phi)(e^{-\Phi}\eta_0 e^\Phi)
- \int_0^1 dt (e^{-\hat\Phi}\p_t e^{\hat\Phi})
\{ e^{-\hat\Phi}Q e^{\hat\Phi} ~,~ e^{-\hat\Phi}\eta_0 e^{\hat\Phi}\}\rangle 
\label{ns}
\end{equation}
where $e^\Phi = 1 + \Phi + \half \Phi * \Phi + ...$ is defined
using the midpoint gluing prescription, 
$\Phi$ is the NS string field discussed earlier, 
$\hat\Phi(t=0) = 0$ and $\hat\Phi(t=1)=\Phi$.
One can show
that $S$ is invariant under the WZW-like gauge invariance
$$\d(e^\Phi) =  e^\Phi (\eta_0\tilde\Lambda) + (Q\Lambda)  e^\Phi, $$
and that on-shell,
$$\eta_0 (e^{-\Phi} Q e^{\Phi}) =0,$$ 
which generalize the linearized gauge invariance and
equations of motion of the quadratic action.

To explicitly evaluate the action of (\ref{ns}),
one first performs a Taylor expansion
in $\Phi$ to obtain
$$S= {1\over{\lambda^2}}
Tr \langle \half \Phi Q\eta_0 \Phi - {1\over 6}\Phi\{Q\Phi,\eta_0\Phi\}
+ ...\rangle$$
where all string fields are multiplied together using the midpoint
interaction.
The cubic term can be evaluated in precisely the same manner as in
the CS-like action by mapping three half circles for the external
states into
$2\pi/3$ wedges in the complex plane using the map
$$f_r^{(3)} = e^{{2\pi i(r-1)}\over 3}
( {{1-iz}\over{1+iz}} )^{2\over 3}$$
for $r=1$ to $3$. 
To define the order $N$ term, one uses
the functions 
$$f_r^{(N)} = e^{{2\pi i(r-1)}\over N}
( {{1-iz}\over{1+iz}} )^{2\over N}$$
for $r=1$ to $N$ 
to map $N$ half circles into $2\pi/N$ wedges in the complex plane.
It has been shown by explicit computation \cite{carlos} that the four-point
tree
amplitude is correctly reproduced by the action of (\ref{ns}), 
and there are indirect arguments
based on gauge invariance that all $N$-point tree amplitudes are correctly
reproduced by this action.
The action of (\ref{ns})
has also been used to compute the NS tachyon potential
\cite{potential}\cite{zw}\cite{swed}\cite{il}\cite{david} and kink solutions
\cite{zw}\cite{ohmtwo}
using the level truncation scheme of \cite{sam}\cite{zs}, 
and the results appear to agree with
the predictions of Sen coming from $D$-brane analysis \cite{senuniv}. 

\section{Open N=2 String Field Theory}

Although the action of the previous section
is manifestly Lorentz invariant, it
is not clear how to generalize it to include the Ramond sector.
At the moment, the only action which includes all sectors of the open
superstring is based on a hybrid formalism of the superstring with
$\hat c=2$ N=2 superconformal invariance. In order to understand
the relation of this hybrid action with that of the previous section, it
will be useful to first recall the relation of the bosonic open string
field theory action and Chern-Simons theory. 

As shown by Witten in \cite{cs}, the
action for open bosonic string field theory,
\begin{equation}
S= {1\over {\lambda^2}}
Tr \langle \half V Q V + {1\over 3} V^3\rangle,
\label{bst}
\end{equation}
can also be used to describe the topological string theory version
of Chern Simons. This Chern-Simons
string theory is defined by a $\hat c=3$
$N=2$ superconformal field theory constructed from the worldsheet
variables $[x^j, \bar x_j, \psi^j,\bar \psi_j]$ for $j=1$ to 3
with the twisted $N=2$ generators:
$$T = \p x^j \p \bar x_j + \bar\psi_j \p \psi^j,$$
$$G^+ = \psi^j \p \bar x_j,\quad G^- = \bar\psi_j \p x^j,$$
$$J = \psi^j \bar\psi_j.$$

If one identifies $Q$ with $\int G^+$ and 
the ghost-number with U(1) charge, 
the action of (\ref{bst}) 
reproduces the Chern-Simon action
$$S_{CS}= {1\over {\lambda^2}}
Tr \int d^3 x \epsilon^{jkl}[\half A_j \p_k A_l + {1\over 3} 
A_j A_k A_l].$$
To obtain the Chern-Simons action from (\ref{bst}), 
one uses the normalization that
$\langle \psi^j\psi^k\psi^l\rangle = \epsilon^{jkl}$ and expands the
ghost-number $+1$ string field $V$ as 
$$V = A_j(x)\psi^j + ... ~~.$$ 
The terms $...$ involve derivatives on $\psi$ and $x$ and describe
massive fields which do not propagate.
One can easily generalize the above construction to any $\hat c=3$
N=2 superconformal field theory and the corresponding action computes
topological quantities in the $N=2$ superconformal field theory \cite{kod}.

The action of (\ref{ns})
in the previous section involves operators $Q$ and $\eta_0$,
but $\eta_0$ has no analog in $\hat c=3$ $N=2$ superconformal field theory.
However, as will now be shown, $Q$ and $\eta_0$ do have analogs
in $\hat c=2$ $N=2$ superconformal field theory, i.e. in critical $N=2$
strings. When $\hat c=2$, the operators $J=\p H$, $J^{++}= e^{H}$ and 
$J^{--} = e^{- H}$ generate an SU(2) affine Lie algebra.
Commuting these SU(2) generators with the fermionic generators $G^+$
and $G^-$ generate two new fermionic generators defined as
$$\tilde G^+ = [\int e^H, G^-],\quad
\tilde G^-= [\int e^{-H}, G^+].$$
These four fermionic generators combine with the SU(2) generators
and the stress tensor to form a set of ``small''
N=4 superconformal generators.
After twisting by $J$, $(G^+,\tilde G^+)$ carry conformal
weight $+1$ and $(G^-,\tilde G^-)$ carry conformal weight $+2$.
So if the U(1) charge is identified with ghost-number, one can
identify $\int G^+$ with $Q$ and $\int \tilde G^+$ with $\eta_0$ \cite{top}.

There are three critical N=2 superconformal field theories which
will be relevant here. The first is the self-dual string which describes
self-dual Yang-Mills \cite
{oog}. The worldsheet variables of the self-dual string consist of 
$[x^j,\bar x_j,\psi^j,\bar\psi_j]$ where $j=1$ to 2.
For the self-dual string, the 
N=4 generators after twisting are 
$$T = \p x^j \p\bar x_j + \bar\psi_j \p\psi^j,$$
\begin{equation}
G^+ = \psi^j \p\bar x_j,\quad \tilde G^+ = \epsilon_{jk}\psi^j \p x^k ,
\label{selfdual}
\end{equation}
$$G^- =\bar\psi_j \p x^j,\quad
 \tilde G^- = \epsilon^{jk} \bar\psi_j \p \bar x_k, $$
$$J^{++}= \half\e_{jk}\psi^j\psi^k,\quad J = \psi^j\bar\psi_j,\quad
J^{--}=\half\e^{jk}\bar\psi_j\bar\psi_k.$$

If one replaces $Q$ with $\int G^+$ and $\eta_0$ with $\int \tilde G^+$ of
(\ref{selfdual}),
the action of (\ref{ns})
reproduces the Donaldson-Nair-Schiff action \cite
{don}\cite{nair} for
self-dual Yang-Mills,
$$S = {1\over {2\lambda^2}}
Tr\int d^4 x ((e^{-\phi} \p_j e^\phi)(e^{-\phi}\bar\p^j e^\phi)
+ \int_0^1 dt (e^{-\hat\phi}\p_t e^{\hat\phi})[
 e^{-\hat\phi}\p_j e^{\hat\phi}~,~
 e^{-\hat\phi}\bar\p^j e^{\hat\phi}])$$
in terms of the Yang field $\phi(x)$.
To obtain this action from (\ref{ns}), one 
uses the normalization that
$\langle \psi^j\psi^k\rangle = \e^{jk}$ and expands the
ghost-number zero string field $\Phi$ as 
$$\Phi= \phi(x) + ...$$ 
where the terms $...$ involve derivatives on $\psi$ and $x$ and correspond
to non-propagating massive fields.
Note that the equation of motion 
$$\int\tilde G^+[e^{-\Phi} (\int G^+) e^{\Phi}]=0$$
implies 
$\bar\p^j(e^{-\phi}\p_j e^{\phi}) =0 $
which is Yang's equation for self-dual Yang-Mills.

A second critical N=2 superconformal field theory is given by the
N=2 embedding of the RNS superstring. As shown in 
\cite{UST}, any critical N=1 superconformal
field theory (such as the ten-dimensional
superstring) can be described by a critical
N=2 superconformal field theory. 
The worldsheet fields of this critical N=2 superconformal field theory are
the usual RNS worldsheet variables $[x^\mu,\psi^\mu,b,c,\xi,\eta,\phi]$
for $\mu=0$ to 9 and the twisted N=4
generators are defined by \cite{twisted}\cite{UST}\cite{top}
$$T= T_{RNS},$$
\begin{equation}
G^+ = j_{BRST},\quad \tilde G^+ = \eta,
\label{RNS}
\end{equation}
$$G^- = b,\quad \tilde G^- = \{Q, b\xi\} = -b Z + \xi T_{RNS} ,$$
$$J^{++}=c\eta,\quad J= bc+\xi\eta,\quad J^{--}= b\xi,$$
where $Q=\int j_{BRST}$ is the standard BRST charge of the N=1 superstring
and $T_{RNS}$ is the sum of the stress tensors for the RNS matter and
ghost variables.
Since $Q=\int G^+$ and $\eta_0=\int \tilde G^+$,  
this explains the relationship of the
action in the previous section with the actions constructed in this section
using the N=4 superconformal generators.

Finally,
a third critical N=2 superconformal field theory is given by a hybrid
version of the 
superstring which describes in $D=4$ superspace the
superstring compactified on a six-dimensional manifold \cite{four}. 
As will be reviewed in the following section, the open superstring
field theory action \cite{sft}
constructed from this N=2 superconformal field theory
is manifestly $D=4$ super-Poincar\'e covariant and includes all sectors
of the superstring.

\section{Open Superstring Field Theory}

For any compactification of the open superstring to four dimensions
which preserves at least N=1 $D=4$ spacetime supersymmetry, there
exists a field redefinition that maps the RNS worldsheet variables to
a set of hybrid variables which include the four-dimensional Green-Schwarz
variables $[x^m,\t^\a,\bar\t^\ad]$ for $m=0$ to 3 and $\a,\ad=1$ to 2.
The complete set of hybrid variables is given by
$[x^m, \theta^\alpha,$
$ \bar\theta^{\dot\alpha},$
$ p_\alpha,$
$\bar p_{\dot\alpha},$
$ \rho]$ 
plus a twisted N=2 c=9 superconformal field theory
which describes the compactification manifold. $[p_\alpha,\bar p_{\dot\alpha}]$
are the conjugate momenta to the superspace
variables $[\theta^\alpha,\bar\theta^{\dot\alpha}]$ satisfying
the OPE's $p_\a (y) \t^\b(z) \to \d_\a^\b (y-z)^{-1}$ and
$\bar p_\ad (y) \bar\t^\bd(z) \to \d_\ad^\bd (y-z)^{-1}$, and
$\rho$
is a chiral boson with background charge $-1$ satisfying the OPE
$\rho(y)\rho(z) \to -\log(y-z)$. 

The field redefinition from RNS to hybrid variables maps the twisted
N=4 superconformal generators of (\ref{RNS})
to the manifestly $D=4$ super-Poincar\'e covariant generators 
$$T= -\half \p x^m \p x_m - p_\a \p\theta^\a -\bar p_\ad \p \bar\t^\ad
-\half\p\rho\p\rho - \half\p^2\rho + T_C,$$
\begin {equation}
G^+ = d^\alpha d_{\alpha} e^\rho + G^+_C, \quad
\tilde G^+ = [\int e^{-\rho+H_C}, G^-],
\label{GS}
\end {equation}
$$G^- = \bar d^{\dot\alpha} \bar d_{\dot\alpha} e^{-\rho} + G^-_C,\quad
\tilde G^- = [\int e^{\rho-H_C}, G^+],$$
$$J^{++}= e^{-\rho+ H_C},\quad J=-\p\rho+ \p H_C,\quad
J^{--}= e^{\rho- H_C},$$
where 
$$d_\alpha=p_\alpha +{i\over 2}
\bar\theta^{\dot\alpha}\s^m_{\alpha\dot\alpha} \p x_m
-{1\over 4} (\bar\t)^2 \p\t_\a +{1\over 8} \t_\a \p(\bar\t)^2,$$
$$\bar d_{\dot\alpha}=\bar p_{\dot\alpha} +{i\over 2}\theta^{\alpha}
\s^m_{\alpha\dot\alpha} \p x_m
-{1\over 4} (\t)^2 \p\bar\t_\ad +{1\over 8} \bar\t_\ad \p(\t)^2$$
are spacetime
supersymmetric combinations of the fermionic momenta
and $[T_C, G^+_C, G^-_C, J_C=\p H_C]$ are the twisted
c=9 N=2 superconformal generators
representing the compactification.

In the RNS formalism, one needs to choose a picture for each off-shell
state in the string field theory action. As discussed in \cite{sft},
this choice is replaced in the hybrid formalism by restricting the $\rho$
charge for any off-shell state to be $(-1,0,+1)$. 
So the U(1)-neutral off-shell string field $\Phi$
can be written in an N=1 $D=4$ super-Poincar\'e invariant manner as 
$$\Phi=  \Phi_{-1} +\Phi_0 +  \Phi_{1}$$ 
where $\Phi_n$ are string fields carrying $n$ units of $\rho$ charge
and $-n$ units of
Calabi-Yau U(1) charge. As will be seen later,
the four-dimensional super-Yang-Mills multiplet is contained in $\Phi_0$
and the Calabi-Yau chiral and anti-chiral moduli are contained in
$\Phi_{1}$ and $\Phi_{-1}$.

Because of the different amounts of $\rho$ charge in $G^+$ and
$\tilde G^+$, the linearized
gauge invariance
$\d \Phi= \int G^+\L +\int\tilde G^+
\tilde \L$ generalizes in the hybrid formalism to
\begin{equation}
\delta\P_{-1}=\Gpf\L_{-2} +\Gps\L_{-1}+ \Gtps\L_{0}+\Gtpf\L_1,
\label{gaugelin}
\end{equation}
$$\delta\P_{0}=\Gpf\L_{-1} +\Gps\L_{0}+ \Gtps\L_{1}+\Gtpf\L_2,$$
$$\delta\P_{1}=\Gpf\L_{0} +\Gps\L_{1}+ \Gtps\L_{2}+\Gtpf\L_3$$
where 
$$G_1^+ = \int d^\a d_\a e^\rho, \quad G_0^+ = \int G_C^+,$$
$$ \tilde G_{-1}^+ =\int [\int e^{-\rho +H_C}, G_C^-], \quad
\tilde G_{-2}^+ = \int e^{-2\rho+H_C}
\bar d^\ad \bar d_\ad,$$
and $\L_n$ are gauge parameters carrying $n$ units of $\rho$ charge
and $-1-n$ units of Calabi-Yau charge. 
Note that the cohomology of $\Gtpf$ and $\Gpf$ is trivial since
$\Gtpf({1\over 4}e^{2\rho- H_C}
\bar\t^\ad \bar\t_\ad)=1$
and $\Gpf({1\over 4}e^{-\rho}
\t^\a \t_\a)=1$.
So
the gauge transformations generated by $\L_3$ and $\L_{-2}$ allow
one to gauge-fix 
\begin{equation}
\P_1
=({1\over 4}e^{2\rho-H_C}
\bar\t^\ad \bar\t_\ad) \Omega_{-1}, \quad 
\P_{-1}
=({1\over 4}e^{-\rho}
\t^\a \t_\a)\Omega_0
\label{gf}
\end{equation}
$${\rm where} \quad
\Omega_{-1}=\Gtpf\Phi_1,\quad \Omega_0 = \Gpf\Phi_{-1}.$$

One can easily check that the linearized equations of motion
\begin{equation}
\Gtpf\Gpf\P_{-1}+\Gtpf\Gps\P_0+\Gtpf\Gtps\P_1=0,
\label{eom}
\end{equation}
$$(\Gtps\Gps+\Gtpf\Gpf)\P_{0}+\Gtpf\Gps\P_1+\Gtps\Gpf\P_{-1}=0,$$
$$\Gpf\Gps\P_{-1}+\Gpf\Gtps\P_0+\Gpf\Gtpf\P_1=0,$$
are invariant under the linearized gauge invariances of (\ref{gaugelin})
and therefore generalize the $\int G^+\int \tilde G^+\Phi=0$ equation
in the hybrid formalism.
It was shown in \cite{sft}
that the cohomology of these equations of motion up to
the gauge invariances
of (\ref{gaugelin})
correctly reproduces the RNS cohomology.

To find the non-linear version of these equations of motion and
gauge invariances, it will be useful to see how (\ref{eom})
and (\ref{gaugelin})
describe the massless sector of the uncompactified superstring which
corresponds to $D=10$ super-Yang-Mills. When written in N=1 $D=4$
superspace, $D=10$ super-Yang-Mills is described by the superfields
\begin{equation}
v(x^\mu,\t^\a,\tb^\ad),\quad \oj(x^\mu,\t^\a),\quad
\obj(x^\mu,\tb^\ad)
\label{sf}
\end{equation}
for $\mu=0$ to 9, $(\a,\ad)=1$ to 2, and $j=1$ to 3, 
where $\oj$ and $\obj$ are chiral and anti-chiral 
superfields satisfying $\bar D_\ad\oj = D_\a \obj =0.$
As usual, the N=1 $D=4$ supersymmetric derivatives are defined by 
$D_\a = {\p\over{\p\t^\a}} - {i\over 2} \s^m_{\a\ad} \tb^\ad \p_m$ and
$\bar D_\ad = {\p\over{\p\tb^\ad}} - {i\over 2} \s^m_{\a\ad} \t^\a \p_m$ where
$m=0$ to 3.
The $\t^\a \s^m_{\a\ad}\tb^\ad$ component of $v$ 
describes the four-dimensional polarizations of the Yang-Mills
gauge field while
the $\t$-independent components of $\oj$ and $\obj$ describe the other six
polarizations of the gauge field.

The N=1 $D=4$ superfields of (\ref{sf}) appear in the string field as
$$\Phi_0 = v(x,\t,\tb) + ...,\quad \Omega_{-1}
= \half \epsilon_{jkl} e^{-\rho}
 \psi^j \psi^k \omega^l(x,\t) + ...,\quad
\Omega_0 = \psi^j\obj(x,\tb) + ...,$$
where $\psi^j$ and $\bar\psi_j$ are 
the worldsheet fermions in the internal directions.
Using $G_C^+= \psi^j \p\bar x_j$, $G_C^- = \bar\psi_j \p x^j$ and
$J_C = \psi^j\bar\psi_j$
where $x^j= x_{3+j}+i x_{6+j}$ and
$\bar x_j= x_{3+j}-i x_{6+j}$,
one can check
that the linearized equations of motion 
of (\ref{eom}) imply that 
\begin{equation} 
\bar D^2 \obj + \p_j \bar D^2 v + \epsilon_{jkl} \bar\p^k \omega^l=0,
\label{desired}
\end{equation}
$$(\p_j \bar\p^j + D^\a \bar D^2 D_\a) v + \p_j \omega^j + \bar\p^j\obj =0,$$
$$ D^2 \omega^j + \bar\p^j D^2 v + \epsilon^{jkl} \p_k \bar\omega_l=0,$$
which are the desired equations for these massless superfields.
These equations are invariant under the linearized gauge transformations
\begin{equation}
\d v= D^2 s + \bar D^2 \bar s ,\quad
\d \omega^j=  -\bar\p^j \bar D^2\bar s ,\quad
\d \obj=  -\p_j D^2 s ,
\label{inva}
\end{equation}
which come from choosing $\Lambda_{-1} = e^{-\rho} s(x,\t,\tb)$
and $\Lambda_2 = e^{2\rho -H_C} \bar s(x,\t,\tb)$ in (\ref{gaugelin}).

The non-linear versions of (\ref{desired}) and (\ref{inva}) can
be obtained by covariantizing the  
four-dimensional
superspace derivatives and six-dimensional spacetime derivatives as
\cite{marcus}
\begin{equation}
\N_\a=e^{-v} D_\a e^v,\quad \Nb_\ad=\bar D_\ad,\quad
\N_j= e^{-v}(\p_j+\obj)e^v,\quad \bar\N^j= \bar\partial^j -\oj.
\label{covdev}
\end{equation}
These covariant derivatives satisfy the identities
\begin{equation}
F_{\a\beta}=F_{\ad\dot\beta}=F_{\a j}=F_\ad^j=0
\label{identities}
\end{equation}
where $F_{AB}=\{\N_A, \N_B]$, and transform as
$\delta\N_A= [\N_A, \s]$ under the gauge-transformation
$$\delta e^v=\bar \s e^v + e^v \s,\quad
\delta\oj= -\bar\partial^j  \s +[\oj, \s],\quad
\delta\obj= -\p_j \bar \s -[\obj, \bar \s]$$
where $\bar\s= D^2 s$ and $\s = \bar D^2 \bar s$ for arbitrary
$s(x,\t,\tb)$ and $\bar s(x,\t,\tb)$.

In terms of these field strengths, the non-linear
equations of motion for ten-dimensional super-Yang-Mills are
\begin{equation}
2\{\N^\a, W_\a\}=F_j^j, \quad
2\{\N^\a, F_\a^j\}=\epsilon^{jkl} F_{kl},\quad
2\{\Nb^\ad, F_{\ad j}\}=\epsilon_{jkl} F^{kl},
\label{onshell}
\end{equation}
where $W_\a=[\Nb^\ad,\{\N_\a,\Nb_\ad\}]
= \bar D_\ad \bar D^\ad (e^{-v} D_\a e^v)$
is the four-dimensional
chiral field strength. The action which produces these
equations of motion is \cite{marcus}
\begin{equation}
S=\half\int d^{10}x [\,\, -2\int d^2 \t \, W^\a W_\a  
\label{superpart}
\end{equation}
$$+
\int d^4 \t \left( (e^{-v}\pj e^v)(e^{-v}\pjb e^v)-\int_0^1 dt
(e^{-\hat v}\partial_t e^{\hat v})
\{ e^{-\hat v}\pj e^{\hat v},
e^{-\hat v}\pjb e^{\hat v}\} )\right)
$$
$$+2\int d^4\t\left( (\pjb e^{-v})\obj e^v+
 e^{v}\oj(\pj e^{-v})+ e^{-v}\obj e^v\oj\right)$$
$$ +\int d^2\t \epsilon_{jkl}
(\oj\bar\partial^k\omega^l+{2\over 3}\oj\omega^k\omega^l)+
\int d^2\tb \epsilon^{jkl}
(\obj\partial_k\obl-{2\over 3}\obj\obk\obl)\,\,].$$

Using intuition from the point-particle example, it is now
straightforward to guess the non-linear versions of the superstring
equations of motion and gauge invariances of (\ref{eom})
and (\ref{gaugelin}). In analogy with (\ref{covdev}),
one first defines the covariantized operators
\begin{equation}
\Gf=e^{-\Phi_0}\Gpf e^{\Phi_0},\quad \Gtf=\Gtpf, 
\label{stringcov}
\end{equation}
$$
\Gs=e^{-\Phi_0}(\Gps+\Omega_0)e^{\Phi_0}, \quad\Gts=\Gtps-\Omega_{-1},$$
where $\Ob\equiv\Gpf\P_{-1}$ and $\O\equiv
\Gtpf\P_1$.
Like their point-particle counterparts in (\ref{covdev})
and (\ref{identities}), 
these covariantized operators
satisfy the identities
\begin{equation}
\{\Gf,\Gf\}=\{\Gtf,\Gtf\}=\{\Gf,\Gs\}=\{\Gtf,\Gts\}=0
\label{stringid}
\end{equation}
and transform as $\delta{\cal{G}}_A=[{\cal{G}}_A,\Sigma]$ under
the gauge transformations
\begin{equation}
\delta e^{\Phi_0}=\bar\Sigma e^{\Phi_0} +e^{\Phi_0} \Sigma,\quad
\delta\O=-\Gtps\Sigma
+[\O,\Sigma],\quad
\delta\Ob=-\Gps\bar\Sigma-[\Ob,\bar\Sigma]
\label{sg}
\end{equation}
where $\bar\Sigma=\Gpf\L_{-1}$ and $\Sigma=\Gtpf\L_2$.

A natural string generalization of the point-particle equations of
motion in equation (\ref{onshell}) is
\begin{equation}
\{\Gf,\Gtf\}=-\{\Gs,\Gts\},
\end{equation}
$$2\{\Gf,\Gts\}=-\{\Gs,\Gs\},\quad
2\{\Gtf,\Gs\}=-\{\Gts,\Gts\}.$$
These equations can be combined with the identities of (\ref{stringid})
to imply that
\begin{equation}
(\Gf+\Gs+\Gts+\Gtf)^2=0,
\label{ff}
\end{equation}
which is the natural generalization of $(Q+A)^2=0$ for the
Chern-Simons-like action.

In addition to the gauge invariances of equation (\ref{sg}), the equations
of motion implied by (\ref{ff}) are also invariant under 
\begin{equation}
\delta e^{\Phi_0}=e^{\Phi_0} (\Gs\L_0 +\Gts\L_1),
\label{add}
\end{equation}
$$\delta\O=\Gtpf(\Gf\L_0 +\Gs\L_1),\quad
\delta\Ob=\Gpf(e^{\Phi_0}( \Gts\L_0 +\Gtf\L_1)e^{-\Phi_0}).$$
Unlike the gauge transformations of (\ref{sg}),
these gauge transformations have no super-Yang-Mills counterpart since
there is no massless contribution to $\L_0$ or $\L_1$.

Starting from the point-particle
action of (\ref{superpart}), 
one can guess that the open superstring field theory
action is
\begin{equation}
S = {1\over {2\lambda^2}}
Tr\langle 
\label{supers}
\end{equation}
$$(e^{-\Phi_0}\Gpf e^{\Phi_0})(e^{-\Phi_0}\Gtpf e^{\Phi_0})
-\int_0^1 dt
(e^{-\hat\Phi_0}\partial_t e^{\hat\Phi_0})
\{ e^{-\hat\Phi_0}\Gpf e^{\hat\Phi_0},
e^{-\hat\Phi_0}\Gtpf e^{\hat\Phi_0} \} $$
$$+(e^{-{\Phi_0}}\Gps e^{\Phi_0})(e^{-{\Phi_0}}\Gtps e^{\Phi_0})-\int_0^1 dt
(e^{-\hat\Phi_0}\partial_t e^{\hat\Phi_0})
\{ e^{-\hat\Phi_0}\Gps e^{\hat\Phi_0},
e^{-\hat\Phi_0}\Gtps e^{\hat\Phi_0}\} $$
$$+
2\left( (\Gtps e^{-{\Phi_0}})\Ob e^{\Phi_0}+
 e^{{\Phi_0}}\O(\Gps e^{-{\Phi_0}})+ e^{-{\Phi_0}}\Ob e^{\Phi_0}\O\right)$$
$$-(\O\Gtps\P_{1}-{2\over 3}\O\O\P_{1})+
(\Ob\Gps\P_{-1}+{2\over 3}\Ob\Ob\P_{-1})\,\,\rangle$$
using the normalization definition that
$$\langle (\t)^2 (\tb)^2 e^{-\rho} \psi^j \psi^k\psi^l\rangle = 
\epsilon^{jkl}.$$

To show that the superstring field theory action of
(\ref{supers}) is correct, one can
check that its linearized equations of motion and gauge invariances
reproduce the on-shell conditions of (\ref{eom}) and
(\ref{gaugelin}) for physical vertex operators, and that
the cubic term in the action produces the three-point tree-level
scattering amplitude. 
Note that both the $W^\a W_\a$ and WZW actions for $v$ in
the first two lines of (\ref{superpart}) are 
replaced by WZW actions in
the second and third lines of (\ref{supers}).
Note also that
the chiral and anti-chiral
$F$-terms in the last line of 
(\ref{superpart}) are replaced by the two terms in the last line
of (\ref{supers}).
Using the gauge invariance $\d\Phi_1 = \Gtpf \L_3$ and $\d\Phi_{-1}=
\Gpf\L_{-2}$, one can choose the gauge of (\ref{gf})
and write these ``$D$-terms'' as the chiral and anti-chiral ``$F$-terms''
\begin{equation}
Tr \langle -\O\Gtps\O 
+{2\over 3}\O\O\O \rangle_F + Tr \langle
\Ob\Gps\Ob+{2\over 3}\Ob\Ob\Ob\rangle_{\widetilde F}
\label{fterm}
\end{equation}
where one defines the normalization of $\langle~~\rangle_F$ and
$\langle~~\rangle_{\widetilde F}$ by
$$\langle (\t)^2 e^{-3\rho + 2 H_C}\rangle_F={1\over 4},\quad
\langle (\tb)^2 e^{H_C}\rangle_{\widetilde F} ={1\over 4}.$$
Since $\tilde G^+$ is the inverse of the $\xi$ zero mode,
turning $D$-terms into $F$-terms is like going from the large
to the small RNS Hilbert space.
The $F$-terms in (\ref{fterm})  are expected to satisfy non-renormalization
theorems similar to those satisfied by the point-particle $F$-terms in
the last line of (\ref{superpart}).

One can include both the GSO($+$) and GSO($-$) sectors in the action
of (\ref{supers}) by 
allowing $(\t^\a,\tb^\ad)$
and $(p_\a,\bar p_\ad)$ 
to be integer moded in the GSO($+$) sector and
half-integer moded in the GSO($-$) sector. As in the NS action of
\cite{potential}\cite{zw}, one needs to include extra $2\times 2$
matrices on the string fields $(\Phi_{-1},\Phi_0,\Phi_1)$
and on the operators $(G^+_{1},G^+_0,\tilde G^+_{-1},\tilde G^+_{-2})$
to account for the ``wrong'' statistics in the 
GSO($-$) sector \cite{gso}.
Although $N=1$ spacetime supersymmetry is broken after including the
GSO($-$) sector since $(\t^\a,\tb^\ad)$ no longer has zero modes, 
it has been conjectured by Yoneya that the action contains a hidden
N=2 spacetime supersymmetry which is restored after the tachyon
condenses
\cite{Yon}\cite{Ypr}.

\vskip 15pt

{\bf Acknowledgements:} I would like to thank Marcelo Leite, Ashoke Sen,
Warren Siegel, Carlos Tello Echevarria, Cumrun
Vafa, Tamiaki Yoneya
and Barton Zwiebach for their collaboration on parts of this work,
Irina Aref'eva for discussions on her work, Caltech for their hospitality,
and the organizers of the ICTP Latin-American School and Komaba 2000 Workshop
for very enjoyable conferences.
This research was partially supported by
FAPESP grant 99/12763-0, CNPq grant 300256/94-9, and Pronex
grant 66.2002/1998-9, and
was partially conducted during the period the author
was employed by the Clay Mathematics Institute as a CMI Prize Fellow.

%%%%%%%%%%%%%%%%%%%%%%

\end{document}